\documentclass[12pt]{iopart}

\usepackage{dsfont}
\usepackage{tikz}
\usepackage{ulem}
\usepackage{physics}

\usepackage{amsmath}
\usepackage{float}
\usepackage{url}
\usepackage[
    backend=biber,
    style=ieee,
  ]{biblatex}
\usepackage{hyperref}
\hypersetup{
	colorlinks,linkcolor=blue,citecolor=blue,urlcolor=blue}

\begin{document}

\title{The Dance of the Sheared Eigenfunctions}

\author{J. Oliveira-Cony, F. S. S. Rosa, C. Farina}

\address{Instituto de Física, Universidade Federal do Rio de Janeiro}

\author{Reinaldo de Melo e Souza}
\address{Instituto de Física, Universidade Federal Fluminense}

%\ead{customerservices@ioppublishing.org}

%\address{Universidade Federal do Rio de Janeiro}

\vspace{10pt}
\begin{indented}
\item[]January 2026
\end{indented}

\begin{abstract}
In this work, we delve into the theory of sheared potentials in non-relativistic quantum mechanics. After defining what we mean by a family of sheared potentials, we consider these families in two particular but emblematic cases,  the harmonic oscillator and the  symmetric potential well proportional to $|x|$. In both cases, besides determining the spectra, we analyse the impact of the shearing process
on the respective eigenfunctions. The latter analysis is typically left
aside in the literature, but here we show that the sheared eigenfunctions yield insights that allow for a deeper understanding of the main features exhibited by the spectra. Finally, we make a few comments about the connection between the change in the spectra of a given sheared family and the necessary work that must be made by an external agent to implement such a change.
\end{abstract}

%\maketitle

\section{Introduction: sheared potentials in quantum theory}
\label{sec1}

Potential theories are one of the most important ingredients of a physicist toolbox. Most equations in physics are, in fact, schemes of differential equations with distinct potential functions, as 
for instance, the  Laplace  and the Schr\"odinger equations. Naturally, solving these equations in the widest possible generality is a well established goal in many branches of physics. 
%all of physics. 
With this purpose, we consider an interesting and yet underdeveloped method of creating a parametrized family of potentials which are particularly appealing to be explored in the Schr\"odinger  equation, namely, sheared potentials. 

In order to define precisely what we mean by sheared potentials, let us consider one dimensional periodic motions of a particle along a given axis, which is submitted to a conservative force $F(x) = -\mathcal{U}'(x)$ where $\mathcal{U}(x)$ is the associated potential {energy}. For periodic motions this must be a well of some sort, and we assume it to have just one minimum, which we take to be at $x=0$. Let $\mathcal{E}$ denote the mechanical energy measured with respect to the minimum potential energy and also $a(\mathcal{E}),b(\mathcal{E})$ -- where $a(\mathcal{E})<0<b(\mathcal{E})$ -- be the turning points of the motion, determined by the roots of the equation $\mathcal{E}=\mathcal{U}(x)$.

We define a family of sheared potential {(energy)} wells by the set of potentials $\{ \mathcal{U}_\xi\}$, with $\xi$ being a continuous parameter which deforms continuously the original function $\mathcal{U}$, constrained by the following condition
\begin{equation}\label{SCondition}
b_{\xi'}(\mathcal{E}) - a_{\xi'}(\mathcal{E}) = 
b_{\xi}(\mathcal{E}) - a_{\xi} (\mathcal{E})\, ,
\end{equation}
where $a_\xi(\mathcal{E}),b_\xi(\mathcal{E})$ are the turning points of $\mathcal{U}_\xi(x)$ for the mechanical energy $\mathcal{E}$ (analogously for $a_{\xi'}(\mathcal{E}),b_{\xi'}(\mathcal{E})$), for any allowed value of ${\cal E}$. Equation (\ref{SCondition}) defines a family of sheared potentials, and without loss of generality we assume that in the deformation process the minimum value of $\mathcal{U}_{\xi}$ is zero and happens at $x=0$.

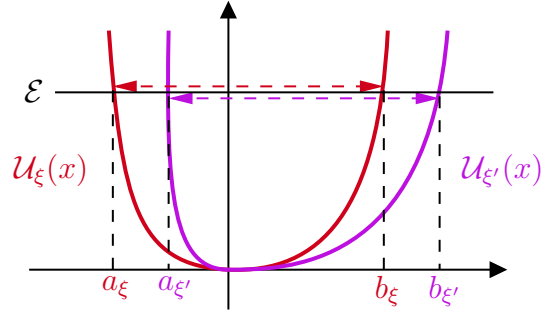
\begin{figure}[H]
\begin{center}

\tikzset{every picture/.style={line width=0.75pt}} %set default line width to 0.75pt        

\begin{tikzpicture}[x=0.75pt,y=0.75pt,yscale=-1,xscale=1]

%Straight Lines [id:da8271468236257218] 
\draw    (180,210) -- (180,58) ;
\draw [shift={(180,55)}, rotate = 90] [fill={rgb, 255:red, 0; green, 0; blue, 0 }  ][line width=0.08]  [draw opacity=0] (8.93,-4.29) -- (0,0) -- (8.93,4.29) -- cycle    ;
%Straight Lines [id:da04120284347764125] 
\draw    (80,190) -- (317,190) ;
\draw [shift={(320,190)}, rotate = 180] [fill={rgb, 255:red, 0; green, 0; blue, 0 }  ][line width=0.08]  [draw opacity=0] (8.93,-4.29) -- (0,0) -- (8.93,4.29) -- cycle    ;
%Curve Lines [id:da6449681068816253] 
\draw [color={rgb, 255:red, 208; green, 2; blue, 27 }  ,draw opacity=1 ][line width=1.5]    (120,70) .. controls (126.33,156) and (133.67,188) .. (180,190) .. controls (226.33,192) and (255.33,161) .. (260,70) ;
%Curve Lines [id:da05338957678474232] 
\draw [color={rgb, 255:red, 189; green, 16; blue, 224 }  ,draw opacity=1 ][line width=1.5]    (150,70) .. controls (148.33,153) and (152.33,190) .. (180,190) .. controls (207.67,190) and (280.33,191) .. (290,70) ;
%Straight Lines [id:da6013601693272637] 
\draw [color={rgb, 255:red, 208; green, 2; blue, 27 }  ,draw opacity=1 ] [dash pattern={on 4.5pt off 4.5pt}]  (124,98) -- (256,98) ;
\draw [shift={(258,98)}, rotate = 180] [fill={rgb, 255:red, 208; green, 2; blue, 27 }  ,fill opacity=1 ][line width=0.08]  [draw opacity=0] (12,-3) -- (0,0) -- (12,3) -- cycle    ;
\draw [shift={(122,98)}, rotate = 0] [fill={rgb, 255:red, 208; green, 2; blue, 27 }  ,fill opacity=1 ][line width=0.08]  [draw opacity=0] (12,-3) -- (0,0) -- (12,3) -- cycle    ;
%Straight Lines [id:da32724038926274757] 
\draw [color={rgb, 255:red, 189; green, 16; blue, 224 }  ,draw opacity=1 ] [dash pattern={on 4.5pt off 4.5pt}]  (152,104) -- (284,104) ;
\draw [shift={(286,104)}, rotate = 180] [fill={rgb, 255:red, 189; green, 16; blue, 224 }  ,fill opacity=1 ][line width=0.08]  [draw opacity=0] (12,-3) -- (0,0) -- (12,3) -- cycle    ;
\draw [shift={(150,104)}, rotate = 0] [fill={rgb, 255:red, 189; green, 16; blue, 224 }  ,fill opacity=1 ][line width=0.08]  [draw opacity=0] (12,-3) -- (0,0) -- (12,3) -- cycle    ;
%Straight Lines [id:da13741250765438373] 
\draw  [dash pattern={on 4.5pt off 4.5pt}]  (122,100) -- (122,190) ;
%Straight Lines [id:da3808404396527312] 
\draw  [dash pattern={on 4.5pt off 4.5pt}]  (150,103) -- (150,190) ;
%Straight Lines [id:da7872899430956037] 
\draw  [dash pattern={on 4.5pt off 4.5pt}]  (258,100) -- (258,190) ;
%Straight Lines [id:da0563269548204266] 
\draw  [dash pattern={on 4.5pt off 4.5pt}]  (286,103) -- (286,190) ;
%Straight Lines [id:da3459163610397673] 
\draw    (93,101) -- (313,101) ;

% Text Node
\draw (115,192) node [anchor=north west][inner sep=0.75pt]  [color={rgb, 255:red, 208; green, 2; blue, 27 }  ,opacity=1 ]  {$a_{\xi }$};
% Text Node
\draw (252,191.4) node [anchor=north west][inner sep=0.75pt]  [color={rgb, 255:red, 208; green, 2; blue, 27 }  ,opacity=1 ]  {$b_{\xi }$};
% Text Node
\draw (143,192) node [anchor=north west][inner sep=0.75pt]  [color={rgb, 255:red, 189; green, 16; blue, 224 }  ,opacity=1 ]  {$a_{\xi' }$};
% Text Node
\draw (279,191.4) node [anchor=north west][inner sep=0.75pt]  [color={rgb, 255:red, 189; green, 16; blue, 224 }  ,opacity=1 ]  {$b_{\xi'}$};
% Text Node
\draw (76,93.4) node [anchor=north west][inner sep=0.75pt]    {$\mathcal{E}$};
% Text Node
\draw (71,130.4) node [anchor=north west][inner sep=0.75pt]  [color={rgb, 255:red, 208; green, 2; blue, 27 }  ,opacity=1 ]  {$\mathcal{U}_{\xi }( x)$};
% Text Node
\draw (296,130.4) node [anchor=north west][inner sep=0.75pt]  [color={rgb, 255:red, 189; green, 16; blue, 224 }  ,opacity=1 ]  {$\mathcal{U}_{\xi' }( x)$};

\end{tikzpicture}
\end{center}
    \caption{Two potential wells belonging to the same family of sheared potentials. Note that the shearing condition, $b_{\xi'} - a_{\xi'} = 
b_{\xi} - a_{\xi}$, is fullfilled.}
\label{Fig1}
\end{figure}

In Fig. \ref{Fig1} we plot two potential wells which belong to the same family of sheared potentials.  The shearing condition given by Eq. (\ref{SCondition}) can be understood graphically in this figure with the aid of the double horizontal arrows.

%A family $\mathcal{U}_\xi(x)$, where $\xi$ is some parameter, is called a sheared family if, for all $E$, the distance of the points $a_{\xi},b_{\xi}$ such that $E=\mathcal{U}_\xi(a_\xi)=\mathcal{U}_\xi(b_\xi)$ is $\xi-$independent:

%

%The shearing condition is, so,
%\begin{equation}
  %  b_{\xi' } -a_{\xi' } =b_{\xi } -a_{\xi }\quad \text{for all }\, \xi,\xi'\,.
%\end{equation}

Classically, families of sheared potential wells are very rich because it can be shown that the periodic motions associated to potential wells belonging to the same family have the same period (as a function of the mechanical energy)  \cite{pippard1989}:

\begin{equation}
\tau_\xi(\mathcal{E})=\sqrt{2m}\int_{a_\xi}^{b_\xi}\dfrac{dx}{\sqrt{\mathcal{E}-\mathcal{U}_\xi(x)}} = 
\sqrt{2m}\int_{a_{\xi'}}^{b_{\xi'}}\dfrac{dx}{\sqrt{\mathcal{E}-\mathcal{U}_{\xi'}(x)}} =
\tau_{\xi'}(\mathcal{E})\, ,
%
%\quad\text{is $\xi-$independent.}
\end{equation}
for all values of $\xi$ and $\xi'$ of the same family. The study of sheared potentials is well established on the classical regime \cite{ carinena2007, brun2008, terra2016, cross2017}, but it is not so widespread in the quantum realm. The investigation of sheared potentials is of interest for the Einstein-Brillouin-Keller quantization scheme and for isocanonical ensembles in statistical mechanics \cite{mamode2010}, as also for the study of Cherkas systems \cite{choudhury2010}, superintegrability of Hamiltonians \cite{feher2025} and the general modern theory of differential equations \cite{gorni2013, ortega2019, bolotin2023}. This re-emerging and fascinating topic is considered, by T. Padmanabhan, one of the \textit{Sleeping Beauties in theoretical physics} \cite{padmanabhan2015}.

Since in quantum mechanics frequency is related to energy, we could expect that the classical isoperiodicity shared by classical sheared potentials would translate into the property of quantum sheared potentials displaying the same energy spectra
\cite{dorignac2005, carinena2006, chouikha2019}. This is correct {when we consider, for example, the Bohr-Wilson-Sommerfeld or the Wentzel–Kramers–Brillouin schemes of quantization, i.e,} at the semi-classical level, but this is not always true in the full quantum theory \cite{katriel1985, carinena1993}. In contrast with semi-classical theory, under shearing the spacing between consecutive energy levels can increase or decrease with the shearing parameter, and the eigenfunctions of Schr\"odinger's equation,

\begin{equation}
    \left[\dfrac{\hbar^2}{2m}\dfrac{\partial^2 }{\partial x^2}-\mathcal{U}_\xi(x)+\mathcal{E}\right]\psi_\xi(x)=0\,,
    \label{shcrodinger}
\end{equation}

\noindent can be substantially altered. 

The process of shearing can be done by a single formula for all monomial wells, i.e., potential energies of the form $\mathcal{U}_n(x)={\kappa}|x|^n$ by the relation 

\begin{equation}
    {\mathcal{U}}_{\nu,n}(x)=\begin{cases}
        \kappa(\nu|x|)^n&\text{when }x\geq 0\,,\\\\
        \kappa(\nu'|x|)^n&\text{when }x< 0
    \end{cases}
    \label{general}
\end{equation}

\noindent where $\kappa$ is a positive constant and $\nu$ and $\nu'$ are positive constants satisfying the shearing condition,
% %
% \begin{equation}
%     \left(\dfrac{1}{\nu}\right)^n+\left(\dfrac{1}{\nu'}\right)^n=2\quad\Longrightarrow\quad \nu'=\left(2-\left(\dfrac{1}{\nu}\right)^{1/n}\right)^{-n}\,.
% \end{equation}
% %
%
%
\begin{equation}
    \dfrac{1}{\nu}+\dfrac{1}{\nu'}=2\quad\Longrightarrow\quad \nu'=\dfrac{\nu}{2\nu-1}\,.
\end{equation}
We take $\nu\in(1/2,1]$ and accordingly $\nu'\in[1,\infty)$.  $\nu\rightarrow 1/2$ is the limit of an impenetrable barrier at the origin and $\nu=1$ is the case of symmetric potential {energy}. This is not the only way of shearing these families of potentials, and in fact there are infinite ways of shearing a particular potential.

In the next sections, we develop a method to study sheared potentials in quantum theory. We find the spectra and the exact eigenfunctions of the first two monomial wells (that is, the linear one and the harmonic oscillator). Usually, the analysis of sheared potentials focuses only in the spectrum of energy. Indeed, the spectra show us that in the full quantum regime isospectrallity among potentials in the same sheared family is broken. In this paper we argue that studying the eigenfunctions allows us to see why this happens. By varying the shearing parameter the eigenfunctions changes continuously. This dance of the sheared eigenfunctions sheds light on what is happening to the spectrum.

\section{Sheared Monomial Well of Order 1}
\label{sec2}

Taking the case $n=1$ on formula (\ref{general}) we get

\begin{equation}
    {\mathcal{U}}_{\nu}(x)=\begin{cases}
        \kappa\nu x&\text{when }x\geq 0\,,\\\\
        -\kappa\left(\dfrac{\nu}{2\nu-1}\right)x&\text{when }x< 0 \label{Un1}
    \end{cases}
\end{equation}

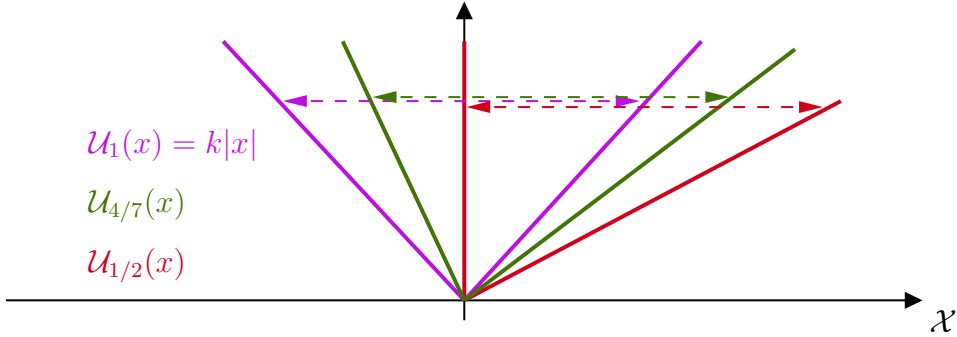
\begin{figure}[H]
\begin{center}

\tikzset{every picture/.style={line width=0.75pt}} %set default line width to 0.75pt        

\begin{tikzpicture}[x=0.75pt,y=0.75pt,yscale=-1,xscale=1]

%Straight Lines [id:da8848359496336629] 
\draw    (221,190) -- (221,33) ;
\draw [shift={(221,30)}, rotate = 90] [fill={rgb, 255:red, 0; green, 0; blue, 0 }  ][line width=0.08]  [draw opacity=0] (8.93,-4.29) -- (0,0) -- (8.93,4.29) -- cycle    ;
%Straight Lines [id:da22876278273810735] 
\draw    (-9,180) -- (448,180) ;
\draw [shift={(451,180)}, rotate = 180] [fill={rgb, 255:red, 0; green, 0; blue, 0 }  ][line width=0.08]  [draw opacity=0] (8.93,-4.29) -- (0,0) -- (8.93,4.29) -- cycle    ;
%Straight Lines [id:da3988958137545766] 
\draw [color={rgb, 255:red, 208; green, 2; blue, 27 }  ,draw opacity=1 ][line width=1.5]    (221,50) -- (221,180) ;
%Straight Lines [id:da7477684497109132] 
\draw [color={rgb, 255:red, 189; green, 16; blue, 224 }  ,draw opacity=1 ][line width=1.5]    (100,50) -- (221,180) ;
%Straight Lines [id:da7140206190435978] 
\draw [color={rgb, 255:red, 189; green, 16; blue, 224 }  ,draw opacity=1 ][line width=1.5]    (221,180) -- (340,50) ;
%Straight Lines [id:da543767507851769] 
\draw [color={rgb, 255:red, 208; green, 2; blue, 27 }  ,draw opacity=1 ][line width=1.5]    (410,80) -- (221,180) ;
%Straight Lines [id:da19653623476131088] 
\draw [color={rgb, 255:red, 65; green, 117; blue, 5 }  ,draw opacity=1 ][line width=1.5]    (160,50) -- (221,180) ;
%Straight Lines [id:da24700729766506613] 
\draw [color={rgb, 255:red, 65; green, 117; blue, 5 }  ,draw opacity=1 ][line width=1.5]    (387,54) -- (221,180) ;
%Straight Lines [id:da879438052704193] 
\draw [color={rgb, 255:red, 189; green, 16; blue, 224 }  ,draw opacity=1 ] [dash pattern={on 4.5pt off 4.5pt}]  (132,80) -- (307,80) ;
\draw [shift={(309,80)}, rotate = 180] [fill={rgb, 255:red, 189; green, 16; blue, 224 }  ,fill opacity=1 ][line width=0.08]  [draw opacity=0] (12,-3) -- (0,0) -- (12,3) -- cycle    ;
\draw [shift={(130,80)}, rotate = 0] [fill={rgb, 255:red, 189; green, 16; blue, 224 }  ,fill opacity=1 ][line width=0.08]  [draw opacity=0] (12,-3) -- (0,0) -- (12,3) -- cycle    ;
%Straight Lines [id:da7586993601775058] 
\draw [color={rgb, 255:red, 65; green, 117; blue, 5 }  ,draw opacity=1 ] [dash pattern={on 4.5pt off 4.5pt}]  (177,78) -- (352,78) ;
\draw [shift={(354,78)}, rotate = 180] [fill={rgb, 255:red, 65; green, 117; blue, 5 }  ,fill opacity=1 ][line width=0.08]  [draw opacity=0] (12,-3) -- (0,0) -- (12,3) -- cycle    ;
\draw [shift={(175,78)}, rotate = 0] [fill={rgb, 255:red, 65; green, 117; blue, 5 }  ,fill opacity=1 ][line width=0.08]  [draw opacity=0] (12,-3) -- (0,0) -- (12,3) -- cycle    ;
%Straight Lines [id:da4395850662030074] 
\draw [color={rgb, 255:red, 208; green, 2; blue, 27 }  ,draw opacity=1 ] [dash pattern={on 4.5pt off 4.5pt}]  (224,83) -- (399,83) ;
\draw [shift={(401,83)}, rotate = 180] [fill={rgb, 255:red, 208; green, 2; blue, 27 }  ,fill opacity=1 ][line width=0.08]  [draw opacity=0] (12,-3) -- (0,0) -- (12,3) -- cycle    ;
\draw [shift={(222,83)}, rotate = 0] [fill={rgb, 255:red, 208; green, 2; blue, 27 }  ,fill opacity=1 ][line width=0.08]  [draw opacity=0] (12,-3) -- (0,0) -- (12,3) -- cycle    ;

% Text Node
\draw (453,183.4) node [anchor=north west][inner sep=0.75pt]    {$\mathcal{X}$};
% Text Node
\draw (31,152.4) node [anchor=north west][inner sep=0.75pt]  [color={rgb, 255:red, 208; green, 2; blue, 27 }  ,opacity=1 ]  {$\mathcal{U}_{1/2}( x)$};
% Text Node
\draw (31,92.4) node [anchor=north west][inner sep=0.75pt]  [color={rgb, 255:red, 189; green, 16; blue, 224 }  ,opacity=1 ]  {$\mathcal{U}_{1}( x) =k|x|$};
% Text Node
\draw (31,122.4) node [anchor=north west][inner sep=0.75pt]  [color={rgb, 255:red, 65; green, 117; blue, 5 }  ,opacity=1 ]  {$\mathcal{U}_{4/7}( x)$};

\end{tikzpicture}
\end{center}
    \caption{Three sheared potentials of the family of the first monomial well. The shearing condition is also indicated in the figure with the horizontal double arrows.}
    \label{Fig2}
\end{figure}

In Fig. \ref{Fig2} we plot three sheared potentials of this family. 

To solve Schr\"odinger's equation with this potential, we suppose a piecewise solution
\begin{equation}
\psi_\nu(x)=\begin{cases}
    \psi_\nu^{(+)}(x)\quad\text{when}\quad x\geq0,\\
    \psi_\nu^{(-)}(x)\quad\text{when}\quad x<0.
\end{cases}
\end{equation}
\noindent which must satisfy the boundary conditions
\begin{align}
    \psi_\nu^{(+)}(0)=\psi_\nu^{(-)}(0)\quad\text{and}\quad
    \dfrac{d\psi_\nu^{(+)}}{dx}(0)=\dfrac{d\psi_\nu^{(-)}}{dx}(0)\,.
    \label{conditions}
\end{align}
For $\nu=1$, we have the symmetric case, and the general solution is a combination of Airy functions \cite{landau1991},
\begin{equation}
    \psi_1{(x)}=\alpha\mathrm{Ai}\left(-\dfrac{E-k |x|}{k^{2/3}}\right)+\beta\mathrm{Bi}\left(-\dfrac{E-k|x|}{k^{2/3}}\right)\,,
\end{equation}

\noindent where we have defined $E=2m\mathcal{E}/\hbar^2$ and $k=2m\mathcal{\kappa}/\hbar^2$. {The term containing $\mathrm{Bi}$ must be discarded, since it diverges for $|x|\to\infty$. For more informations about the Airy functions, see \ref{AppA}.}

It is not difficult to generalize the previous result for an  arbitrary $\nu$. In this case, it is straightforward to show  that $\psi^{(\pm)}_\nu (x)$ are also given by sums of Airy functions,
\begin{align}
    \psi_\nu^{(+)}(x)&=\alpha_+\mathrm{Ai}\left(-\dfrac{E-k\nu x}{(k\nu)^{2/3}}\right)+\beta_+\mathrm{Bi}\left(-\dfrac{E-k\nu x}{(k\nu)^{2/3}}\right)\, ,\\
    \psi_\nu^{(-)}(x)&=\alpha_-\mathrm{Ai}\left(-\dfrac{E+k\nu x/(2\nu-1)}{(k\nu/(2\nu-1))^{2/3}}\right)+\beta_-\mathrm{Bi}\left(-\dfrac{E+k\nu x/(2\nu-1)}{(k\nu/(2\nu-1))^{2/3}}\right)\, .
\end{align}
{Note that this generalization does not change the fact that terms containing $\mathrm{Bi}$ must be discarded}. Using the conditions written in Eqs. \ref{conditions}, we find the following  system of equations for the variables $\alpha_+$ and $\alpha_-$,
\begin{align}
    &\alpha_+\mathrm{Ai}\left(-\dfrac{E}{(k\nu)^{2/3}}\right)=\alpha_-\mathrm{Ai}\left(-\dfrac{E}{(k\nu/(2\nu-1))^{2/3}}\right) \, ,\label{condition1}\\
    &\alpha_+\left(k\nu\right)^{1/3}\mathrm{Ai}'\left(-\dfrac{E}{(k\nu)^{2/3}}\right)=\alpha_-\left(\dfrac{k\nu}{2\nu-1}\right)^{1/3}\mathrm{Ai}'\left(-\dfrac{E}{(k\nu/(2\nu-1))^{2/3}}\right)\, .\label{condition2}
\end{align}
From Eq.(\ref{condition1}), we have
\begin{align}
    \alpha_-=\alpha_+{\mathrm{Ai}\left(-\dfrac{E}{(k\nu)^{2/3}}\right)}{\mathrm{Ai}\left(-\dfrac{E}{(k\nu/(2\nu-1))^{2/3}}\right)}^{-1}\,.
    \label{alphas}
\end{align}
Using the previous equation, we obtain the explicit expressions for the wave functions,
\begin{align}
    \psi_\nu^{(+)}(x) 
&=
\alpha_+\mathrm{Ai}\left(-\dfrac{E-k\nu x}{(k\nu)^{2/3}}\right)\label{psinu+}\\
\psi_\nu^{(-)}(x)
&=
\alpha_+{\mathrm{Ai}\left(\!\!-\dfrac{E}{(k\nu)^{2/3}}\right)}{\mathrm{Ai}\left(\!\!-\dfrac{E}{(k\nu/(2\nu-1))^{2/3}}\right)}^{-1}\!\!\!\mathrm{Ai}\!\left(\!\!-\dfrac{E+k\nu x/(2\nu-1)}{(k\nu/(2\nu-1))^{2/3}}\right),\label{psinu-}
\end{align}
where $\alpha_+$ is a mere normalization constant. 

As it happens, the system formed  by Eqs. (\ref{condition1}) and (\ref{condition2}) is homogeneous, meaning that it has {nontrivial} solutions for $\alpha_{\pm}$ (i.e., solutions allowing $\alpha_{\pm} \neq 0$) if and only if the determinant of the coefficients vanishes, leading to a transcendental equation for $E$. {This allows us to write a function,}
%
%\begin{align}
%\begin{split}
 %   &\left(-\dfrac{1}{2\nu-1}\right)^{1/3}\mathrm{Ai}\left(-\dfrac{E}{(k\nu)^{2/3}}\right)\mathrm{Ai}'\left(-\dfrac{E}{(k\nu/(2\nu-1))^{2/3}}\right)\\
  %  &=\mathrm{Ai}'\left(-\dfrac{E}{(k\nu)^{2/3}}\right)\mathrm{Ai}\left(-\dfrac{E}{(k\nu/(2\nu-1))^{2/3}}\right)\,.
%\end{split}
%\end{align}
%
%For a given value of $\nu$, which means a given potential well of the sheared family under consideration, this equation is satisfied only for specific values of $E$, which are the allowed eigenvalues of the problem. In other words, finding the spectrum of the potential $\mathcal{U}_\nu(x)$ means determining the zeros of the function
%
 \begin{align}
    \begin{split}
        F_\nu(E):&=\left(-\dfrac{1}{2\nu-1}\right)^{1/3}\mathrm{Ai}\left(-\dfrac{E}{(k\nu)^{2/3}}\right)\mathrm{Ai}'\left(-\dfrac{E}{(k\nu/(2\nu-1))^{2/3}}\right)+\\&-\mathrm{Ai}'\left(-\dfrac{E}{(k\nu)^{2/3}}\right)\mathrm{Ai}\left(-\dfrac{E}{(k\nu/(2\nu-1))^{2/3}}\right)\,,
    \end{split}
 \end{align}
\noindent {whose zeros give, for each $\nu$, the spectrum associated with $\mathcal{U}_\nu(x)$.}

\begin{figure}[t]
    \centering
    \includegraphics[width=0.9\linewidth]{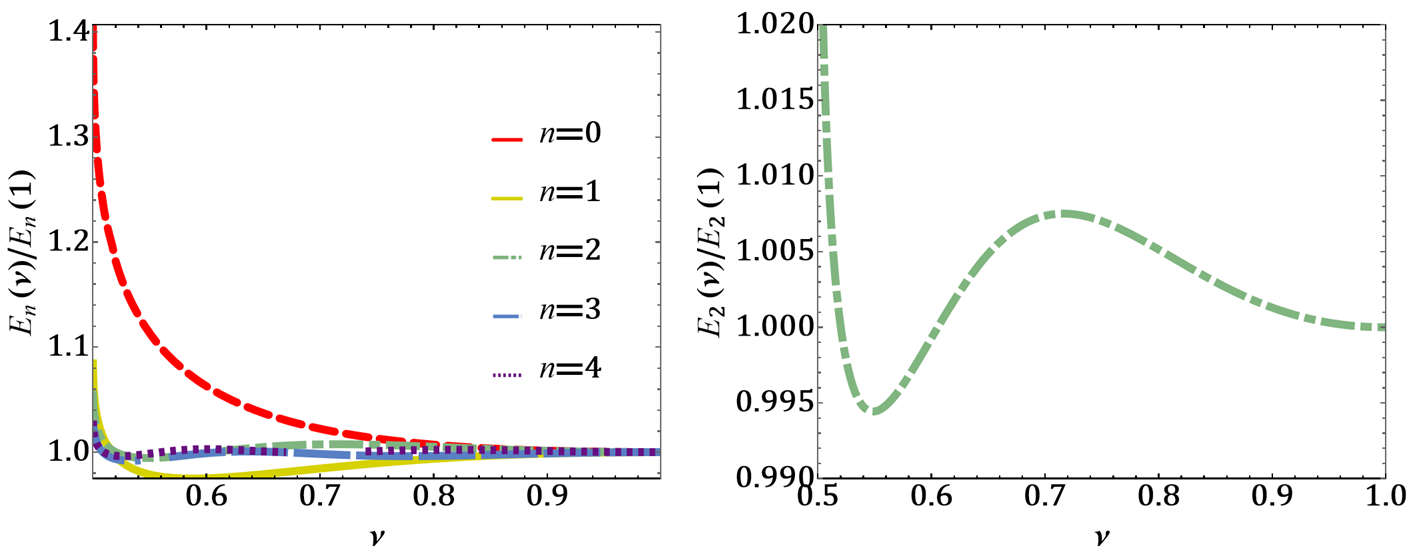}
    \caption{In the left panel we plot the normalized energy levels, $E_n(\nu)/E_n(1)$, for $n=0,1,2,3,4$, in terms of $\nu$. In the right panel we give a zoom in the plot of $E_2(\nu)$ in terms of $\nu$ to emphasize the oscillations for $\nu$ close to $1$.}
    \label{MW1}
\end{figure}

Once we know how to calculate the energy spectrum of any potential well belonging to this family, we can investigate what happens to the energy levels as we vary continuously the parameter $\nu$. {As it is impossible to find analytically the zeros in terms of $\nu$, we will proceed the analysis numerically.} Let us denote by $E_n(\nu)$ the energy level $n$ of the potential $\mathcal{U}_\nu(x)$. In Fig. \ref{MW1} we plot the first five energy levels as functions of $\nu$. In order to better visualize the changes in these energy levels during the shearing process, we plot each $E_n(\nu)$ normalized by $E_n(\nu=1)$ (the corresponding energy value for the symmetric potential of the family). Note that with this normalization, $E_n(\nu)/E_n(1)\rightarrow 1$ as $\nu \rightarrow 1$. From Fig. \ref{MW1} we also see that, the larger the $n$, the smaller the impact {of the shearing process on the energy spectrum}. This was expected, since for large $n$ semi-classical results should hold and, as we have already mentioned, in the semi-classical limit the energy spectrum is $\nu$-independent. For this family, it seems that $n=4$ is already very large.

 It is worth noticing that the functions $E_n(\nu)$ oscillate, as we can see clearly in the right panel of Fig. \ref{MW1}. To understand these oscillations, one can compute $dE_n/d\nu=0$. This calculation can be made with the help of the so-called Hellmann-Feynman theorem \cite{Wyatt1966}, which states that if we have a hamiltonian $\mathcal{H}$ that depends on a parameter $\nu$, so that the eigenfunctions  $\psi_\nu$ are $\nu-$dependent functions with  $E(\nu)$ corresponding the $\nu$-dependent energy eigenvalues, then we can write
\begin{equation}
    \dfrac{dE(\nu)}{d\nu}=\bra{\psi_\nu}\dfrac{d\mathcal{H}_\nu}{d\nu}\ket{\psi_\nu}\,.
\end{equation}

Since, for the cases considered in this section,  $\dfrac{d\mathcal{H}_\nu}{d\nu}=\dfrac{d\mathcal{U}_\nu}{d\nu}$, we have for the eigenfunctions
\begin{equation}
    \dfrac{dE(\nu)}{d\nu}=\bra{\psi_\nu}\dfrac{d\mathcal{U}_\nu}{d\nu}\ket{\psi_\nu}=\dfrac{\kappa}{(2\nu-1)^2}\int_{-\infty}^{0} |\psi_\nu^{(-)}(x)|^2x\,dx+\kappa\int_0^{\infty} |\psi_\nu^{(+)}(x)|^2x\,dx \,,
\end{equation}

\noindent where the wavefunctions are described by Eqs. (\ref{psinu+}-\ref{psinu-}). The Hellmann-Feynman theorem, although simple in its claim, is very impressive as it provides a direct formula to calculate the rate of variation of the energy levels as we change the parameter $\nu$ (in our case, the parameter which controls the shearing process). 
The integrals above can be solved analytically; however, the final equation for $dE/d\nu$ is transcendental -- as $\psi_\nu^{(\pm)}$ depends on $E(\nu)$ -- and not easy to solve even numerically. In a moment, we will show how to interpret $dE_n/d\nu$ physically without the need to calculate this quantity directly. This process can be considered, naturally, for any other family of sheared potentials.

As the title of this paper suggests, it is also very interesting to investigate the shape of the sheared eigenfunctions. The wavefunctions get progressively expelled from the negative half-line as $\nu \to 1/2$, until being totally restricted to the positive half-line as the impenetrable barrier limit is achieved. It is important to notice that the energies of each of these functions are different, as $E$ is $\nu-$dependent. 

\begin{figure}[H]
    \centering
    \includegraphics[width=0.9\linewidth]{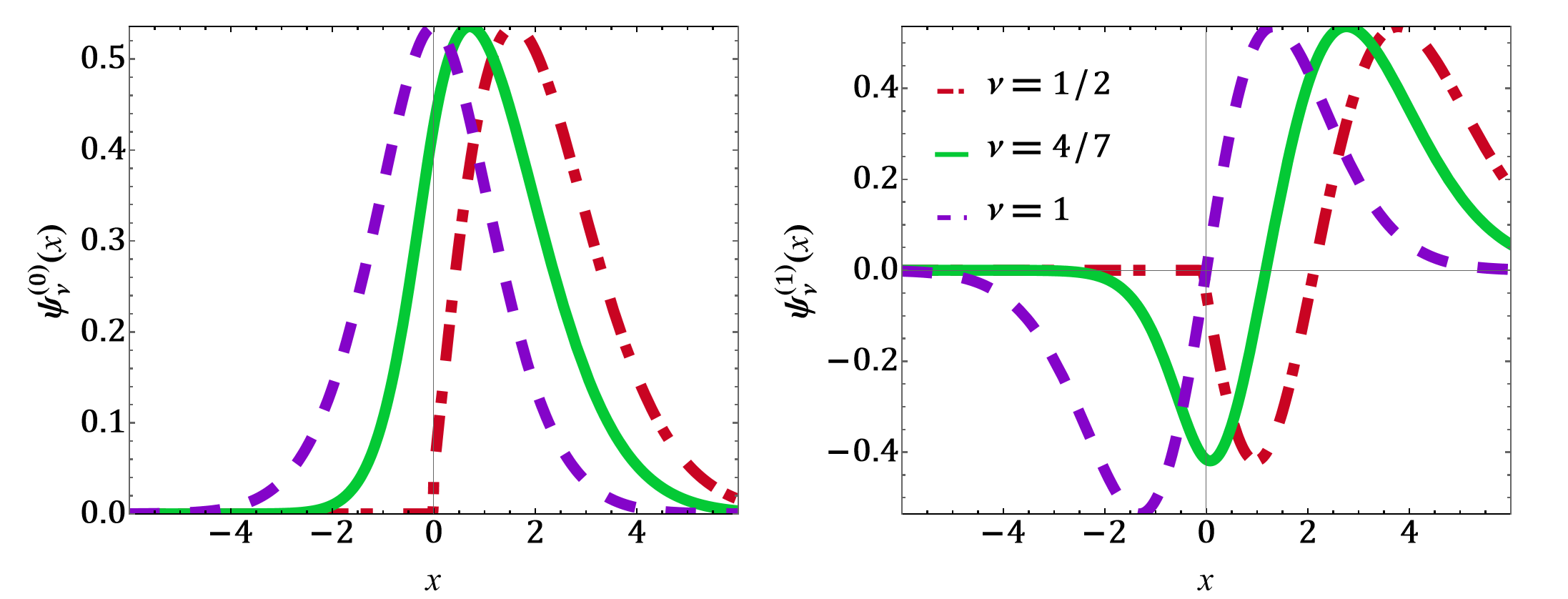}
    \caption{Complete eigenfunctions $\psi_\nu(x)$ for some values of $\nu$ for the first two energy levels (the fundamental state on the left and the first excited state on the right). In these plots, we considered $k=1$.}
    \label{Airypsi}
\end{figure}

We may connect some properties presented by the eigenfunctions with the spectra presented before. If we think of shearing as a physical process by which a given potential gets deformed, then it surely has an energy cost. An external agent is needed to provide the energy for shearing the potential under consideration and this is clearly seen in the $\nu$-dependence of the energy levels. Naturally, if the shearing is conducted abruptly it is well known that the quantum particle can make a transition to a different energy level. In order to analyse the behaviour of the eigenfunction we can think of an experiment where the system is prepared in a given eigenstate and the potential is very gently sheared, so that the particle remains on that state in the course of the whole process. In that picture, the particle would be dragged along with the deformation of the potential, meaning that the wavefunction describing the quantum particle is displaced through space and that the external force is performing work. At this point, we remind the reader that one should exercise a good deal of caution when discussing forces in a quantum mechanical context. For our purposes, we can think of a force as an operator defined, in space representation, by $F(x)=-\mathcal{U}'(x)$.  As can be seen in Eq.~(\ref{Un1}), the force is proportional to the identity operator in this case, and given by $F_+=-\kappa\nu$ in the region $x\geq 0$ and by $F_-=\kappa\nu/(2\nu-1)$ in the region $x<0$, with the signs compatible with the restoring character of the force.

We show now that force can be a useful concept in order to understand the observed decrease of the energy eigenvalues near $\nu=0.5$ in Fig. (\ref{MW1}). The key aspect is to observe that close to $\nu=0.5$ we have $|F_-|\gg |F_+|$.  As we increase $\nu$, the wavefunction enters into the region $x<0$ being shifted to the left (and also deformed). In the process, a negative work is exerted in the region $x<0$ (where the force opposes the shift) while the opposite happens in the region $x>0$. Although the amount of probability that enters region $x<0$ is the same as that which leaves the region of positive $x$, the negative work overcomes the positive one due to the fact that $|F_-|\gg |F_+|$. This explains why $dE/d\nu<0$ for $\nu$ close to $0.5$. When increasing $\nu$, the forces tend to equalize as we are approaching the symmetric case, and thus the  work performed becomes very sensible to the deformation of the wavefunction and to its displacement for every $x$. This explains why the variations of energy are much smaller when $\nu$ is close to $1$ and means that we cannot anticipate the sign of the work and therefore of the change in energy.

\section{Sheared Monomial Well of Order 2 (Harmonic Oscillator)}
\label{sec3}

The shearing of the harmonic oscillator is important due to its deep connection with isochronous potentials \cite{brun2008, cross2017, ortega2019, dorignac2005}, and for the central role that harmonic oscillators have in physics. This problem was already addressed in the context of problems associated to the semi-classical quantization \cite{stillinger1989}, with a similar shearing rule as the one we use in this paper. 
%In their work, the authors analyse the energy variation as the potential is sheared to show some limitations of the semi-classical quantization. 
Here we investigate how the energy varies with the parameter $\nu$ so that we can quantify the breaking of isochronicity, showing the dance of the corresponding eigenfunctions in the process.

Proceeding as the previous case, we solve Schr\"odinger's equation for the potential family
%
% \begin{equation}
%     \mathcal{U}_{\nu}(x)=\begin{cases}
%         \kappa\nu x^2&\text{when }x\geq 0\,,\\\\
%         \kappa{\dfrac{\nu}{(2\sqrt{\nu}-1)^{2}}}x^2&\text{when }x< 0\,.
%     \end{cases}
%     \label{OHshear01}
% \end{equation}
%
%
\begin{equation}
    \mathcal{U}_{\nu}(x)=\begin{cases}
        \kappa\nu^2 x^2&\text{when }x\geq 0\,,\\\\
        \kappa{\dfrac{\nu^2}{(2\nu-1)^2}}x^2&\text{when }x< 0\,.
    \end{cases}
    \label{OHshear01}
\end{equation}
As we are dealing with the harmonic oscillator, is convenient to define $\kappa=m\omega^2/2$. The general solution of Schr\"odinger's equation for the harmonic oscillator is (for $\nu=1$)
\begin{equation}
    \psi(x)=\alpha D_{\frac{\mathcal{E}}{\hbar\omega}-\frac{1}{2}}\left(\left(\dfrac{2m\omega}{\hbar}\right)^{1/2}|x|\right)+\beta D_{-\frac{\mathcal{E}}{\hbar\omega}-\frac{1}{2}}\left(i\left(\dfrac{2m\omega}{\hbar}\right)^{1/2}|x|\right)\,,
    \label{solOH}
\end{equation}
where $D_\sigma(y)$ is the $D-$parabolic cylinder function \cite{whittaker1990}. {For a brief summary about this function, see \ref{AppA}}. When $y$ is purely imaginary, the function diverges for $|y|\to\infty$ so we can henceforth discard the second term in Eq. (\ref{solOH}). For the sheared case, the wave functions are
\begin{align}
\psi_\nu^{(+)}(x)&={\alpha} D_{\sigma(\nu)}\left(k\nu|x|\right)\,,\\
\psi_\nu^{(-)}(x)&={\alpha}\left[\dfrac{D_{\sigma(\nu)}(0)}{D_{\sigma(\mu)}(0)}\right] D_{\sigma(\mu)}\left(k\mu|x|\right)\,,
\end{align}
where $E=\dfrac{\mathcal{E}}{\hbar\omega}$, $k=\sqrt{2m\omega/\hbar}$, 
 $\sigma(\nu)=\dfrac{E}{\nu}-\dfrac{1}{2}$, $\mu=\dfrac{\nu}{2\nu-1}$ and ${\alpha}$ is a normalization constant. Taking the same steps as the development for the linear potential, the transcendental equation for the spectrum in this case reads
\begin{equation}
    (2\nu-1)D_{\sigma(\nu)}'(0)D_{\sigma(\mu)}(0)+D_{\sigma(\nu)}(0)D_{\sigma(\mu)}'(0)\label{trans2}=0\,.
\end{equation}

{In the case $\nu=1$, we have $\mu=1$ and Eq. \ref{trans2} becomes
\begin{equation}
2D_{\sigma(1)}'(0)D_{\sigma(1)}(0)=0\,.
\end{equation}}

{It is possible to show (see \ref{AppA}) that this expression is equivalent to
\begin{equation}
    -\dfrac{2^{-E+1}\pi }{\Gamma\left(-\dfrac{E}{2}+\dfrac{3}{4}\right)\Gamma\left(-\dfrac{E}{2}+\dfrac{1}{4}\right)}=0\,.
\end{equation}}

{As $\Gamma(x)$ only diverges for $(-x)\in\mathds{N}$, the factor $1/\Gamma(-E/2+3/4)$ gives the eigenvalues $E=(2n+1)+1/2$ with $n\in\mathds{N}$ and the factor $1/\Gamma(-E/2+1/4)$ gives the eigenvalues $E=2n+1/2$, also with $n \in \mathds{N}$. As this is merely the decomposition of odd and pair numbers, we can join these results into $E_n=(n+1/2)$ with $n\in\mathds{N}$, as expected. Note that this makes $\sigma(1)\in\mathds{N}$, and in this case the $D-$parabolic cylinder function becomes proportional to a Hermite polynomial, as expected for the traditional harmonic oscillator:
\begin{equation}
    D_n(z)=\mathrm{e}^{-z^2/4}2^{-n/2}H_n(z/\sqrt{2})\,.
\end{equation}}

\begin{figure}[H]
    \centering
    \includegraphics[width=0.9\linewidth]{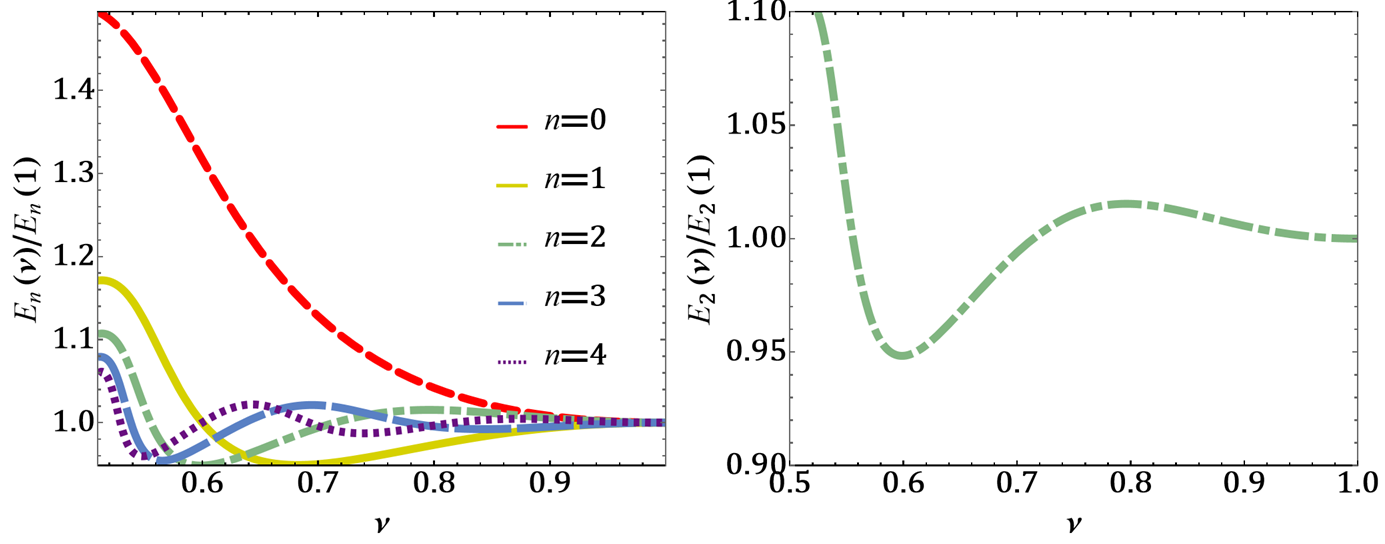}
    \caption{In the left panel we plot the normalized energy levels for the sheared family of the harmonic oscillator, $E_n(\nu)/E_n(1)$, with $n=0,1,2,3,4$, in terms of $\nu$.  In the right panel we give a zoom in the plot of $E_2(\nu)$ to emphasize the oscillations for $\nu$ close to $1$.}
    \label{OHNovoEnergia}
\end{figure}

{For $\nu\neq1$, we are forced, as before, to deal with the problem numerically.} As we vary $\nu$, the first five energy levels change as shown in Figure (\ref{OHNovoEnergia}). Here, we again have normalized our plots by $E_n(1)$ (the energy for the symmetric case). As we have seen in the first monomial well case, the impact of shearing is progressively diminished as $n$ gets larger, which is expected as we are approaching the semi-classical regime.

Now we analyse the behaviour of the eigenfunctions as we shear the potential. Naturally, they must vanish in the region $x<0$ as $\nu\to1/2$. 

\begin{figure}[H]
    \centering
    \includegraphics[width=0.9\linewidth]{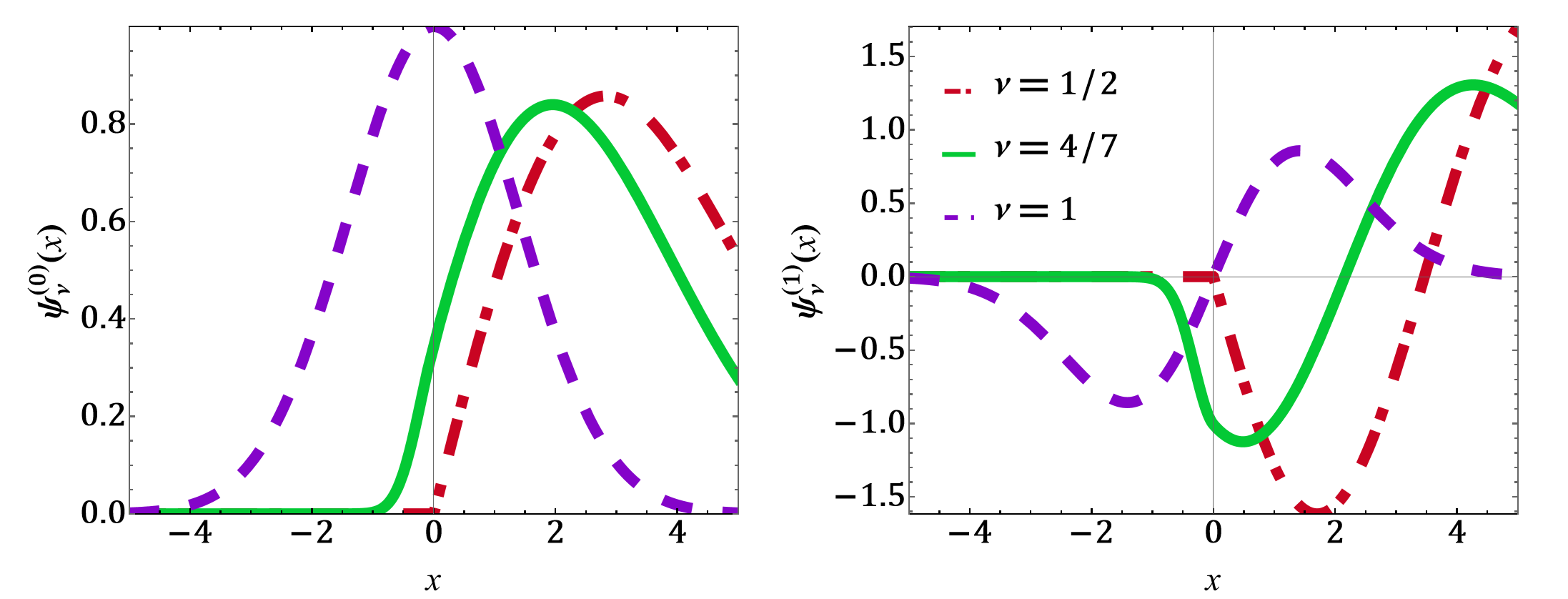}
    \caption{Complete eigenfunctions $\psi_\nu(x)$ for some values of $\nu$ for the first two energy levels (the fundamental state on the left and the first excited state on the right). In these plots, we considered $k=1$.}
    \label{OHpsi1}
\end{figure}

We can again invoke the force concept in order to interpret some features connecting the dance of the eigenfunctions during the shearing process with the changes in the spectrum. The arguments are entirely similar and here we only remark that the relative change in energy is smaller than for the linear potential. This is expected since now the force vanishes at origin, reducing the work while the wavefunctions enter the region $x<0$. 

%This method can be used to treat other interesting potentials, the tricky part being the ability to solve {analytically Schr\"odinger's equation in both positive and negative semi-axis, i.e., to determine $\psi^{(\pm)}(x)$. Of course, one possibility to counter this problem is to take a full numerical route and find the eigenfunctions also with numerical methods.} Once the eigenfunctions are discovered, the sheared spectra and the sheared eigenfunctions can be obtained in the exact same fashion.

It is worth noticing that the eigenfunctions maintain their number of nodes and its oscillatory behaviour for each energy level independently of $\nu$, as it can be seen on Figures (\ref{Airypsi}) and (\ref{OHpsi1}). This is expected by the Sturm oscillation and comparison theorems, which assert that for an appropriate quantum hamiltonian, if $E_1<E_2<\dots$ are the eigenvalues associated with the eigenfunctions $u_1(x,E_1),u_2(x,E_2),\dots$, then $u_n(x,E_n)$ will have exactly $n$ zeros \cite{amrein2005}. The fact that $\psi^{(n)}$ eigenfunctions have exactly $n+1$ nodes is a general feature of Schr\"odinger's equation \cite{moriconi2007}, which is an expression of the more general Courant's nodal domain theorem \cite{Biyikoğu2007}. 

As the eigenfunctions must maintain the number of nodes and the number of zeros independently of $\nu$, the oscillation form of them must be preserved, and the eigenfunctions have to drift as they are sheared. Hence, it is expected that the form of the sheared eigenfunctions must have a similar shape as the ones in the first monomial well, independently of the sheared potential family. Generally, the case $\nu=1/2$ forces a node to appear at $x=0$ due to the infinite barrier at the origin which imposes, by assumption, the boundary condition
 $\psi(0)=0$. 
 Besides, it was recently shown that the oscillation of the functions 
 $E(\nu)$ can be explained by the passage of the zeros 
 of the wavefunctions through $x=0$ (see Ref.\cite{Fernandez} for more details).

\section{Conclusions and final remarks}
\label{conclusion}

In this work, we considered two families of sheared potential wells: a family of linear monomials and a family of harmonic oscillators. In both cases, we have chosen sheared potentials that retained the same functional form as the initial potential for $x < 0$ and $x \ge 0$, but with different coefficients. For both families, we computed not only the spectra but also the corresponding eigenfunctions.  
In order to compute these spectra it was necessary to solve transcendental equations with the aid of numerical methods. According to Felix Klein (1849-1925), ``\textit{it is well known that the central problem of the whole of modern mathematics is the study of transcendental functions defined by differential equations}'' \cite{klein1893}. Although this \textit{modern} is now somewhat old, the pursuit of this kind of treatment, combined with the tools of special functions and modern computing software, is a reviving topic in physics and mathematics on the last few decades \cite{andrews1999special}. 
For the families of sheared potentials considered in this work, we showed that the energy spectrum in both families did not remain constant during the shearing process.

We showed that this is due to the variation of the eigenfunctions. The take-home message of our work is that analysing the sheared eigenfunctions is a key element to understand the change in the spectra. Combining the eigenfunctions with considerations about the work performed we could understand why the spectra display a decrease in energy with increasing $\nu$ in the vicinity of $\nu=0.5$, and also why the variation in this region is more significant then around $\nu=1$. This helps us understand the close similarities displayed by the results for the linear and the harmonic wells and we believe that these features would also be present for other power-law potentials. A perspective of this paper is to deepen this connection, evaluating quantitatively the work done in the process. Once this calculation is made,  one could investigate what would be  the energetic costs during other processes of changing the potential wells in order to find the one which would require a minimum work to link the initial and final potentials.

To investigate how the evolution of quantum systems can be affected by tunable external agents, whose influence are encoded in continuous parameters present in the corresponding hamiltonians, may provide useful mechanisms of controlling the time evolution of such systems The cases considered in this paper are particularly appealing for modelling asymmetric quantum wells, which are used, for instance, on the study of the quantum-confined Stark-effect \cite{Miller1984} or for the Wannier-Stark localization in optical latices \cite{Emin1987}, in the case of the first monomial well. For Stark-related effects, the shearing parameter, which controls the inclination of the curve, is directly proportional to the intensity of the applied electric field. For the harmonic oscillator case, an interesting example is to consider a time-dependent frequency and analyse, with the aid of a parameter that characterizes how abrupt is the change in the frequency, the time evolution of the squeezing parameter for frequencies with different time dependences. One can even investigate a continuous route from an adiabatic change to a completely abrupt one \cite{MartnezTibaduiza2021}. This kind of study may be useful in a rapidly developing area of physics called shortcut to adiabaticity \cite{delCampo2013} and their relation with harmonic traps \cite{Chen2010,Hardel2024}. In this sense, the present work provides some kind of control of quantum spectra and wavefunctions which deserves a deeper investigation, since the shearing process may be useful in a variety of situations.

\section*{Acknowledgments}
\label{ack}

All authors thank M. Asorey for insightful discussions. We also thank the Brazilian agencies CAPES, CNPq and FAPERJ for financial support. J.O.-C. is grateful to FAPERJ (Master's scholarship No. 201.879/2025). C.F. acknowledges funding from CNPq (Grants No. 308641/2022-1 and 408735/2023-6) and FAPERJ (Grant No. 204.376/2024). F.S.S.R. is grateful to CNPq (Grant No. 315730/2021-8).

\appendix

\section{{Résumé of some special functions}}
\label{AppA}

In this paper, we made use of two important special functions, namely the Airy and the $D-$parabolic cylinder functions. In this appendix, we discuss the principal features of these functions.

\subsection{Airy functions}

The Airy functions originate from the Airy (or Stokes) equation,
\begin{equation}
    \dfrac{d^2f}{dz^2}(z)-zf(z)=0\,.
\end{equation}

The general solution of this problem is a sum of the functions called Airy functions, named after the astronomer G. B. Airy, defined by
\begin{align}
    \mathrm{Ai}(z)&=\dfrac{1}{\pi}\int_0^\infty\cos\left(\dfrac{t^3}{3}+zt\right)\,dt\,,\\
    \mathrm{Bi}(z)&=\dfrac{1}{\pi}\int_0^\infty\left\{ \exp\left(-\dfrac{t^3}{3}+zt\right)+\sin\left(\dfrac{t^3}{3}+zt\right)\right\}\,dt\,.
\end{align}

Note that as $\mathrm{Bi}$ diverges for $z\to \infty$, its contribution was discarded in Sec. \ref{sec2}. For analytical calculations, it is useful to write these functions in terms of Bessel functions (consider $x\geq 0$):
\begin{align}
\mathrm{Ai}(x) &= \frac{1}{\pi} \sqrt{\frac{x}{3}} K_{1/3}\left(\frac{2}{3}x^{3/2}\right)\\
\mathrm{Ai}(-x) &= \sqrt{\frac{x}{9}} \left[ J_{1/3}\left(\frac{2}{3}x^{3/2}\right) + J_{-1/3}\left(\frac{2}{3}x^{3/2}\right) \right]\\
\mathrm{Bi}(x) &= \sqrt{\frac{x}{3}} \left[ I_{1/3}\left(\frac{2}{3}x^{3/2}\right) + I_{-1/3}\left(\frac{2}{3}x^{3/2}\right) \right]\\
\mathrm{Bi}(-x) &= \sqrt{\frac{x}{3}} \left[ J_{-1/3}\left(\frac{2}{3}x^{3/2}\right) - J_{1/3}\left(\frac{2}{3}x^{3/2}\right) \right]
\end{align}

\begin{figure}[H]
    \centering
    \includegraphics[width=0.75\linewidth]{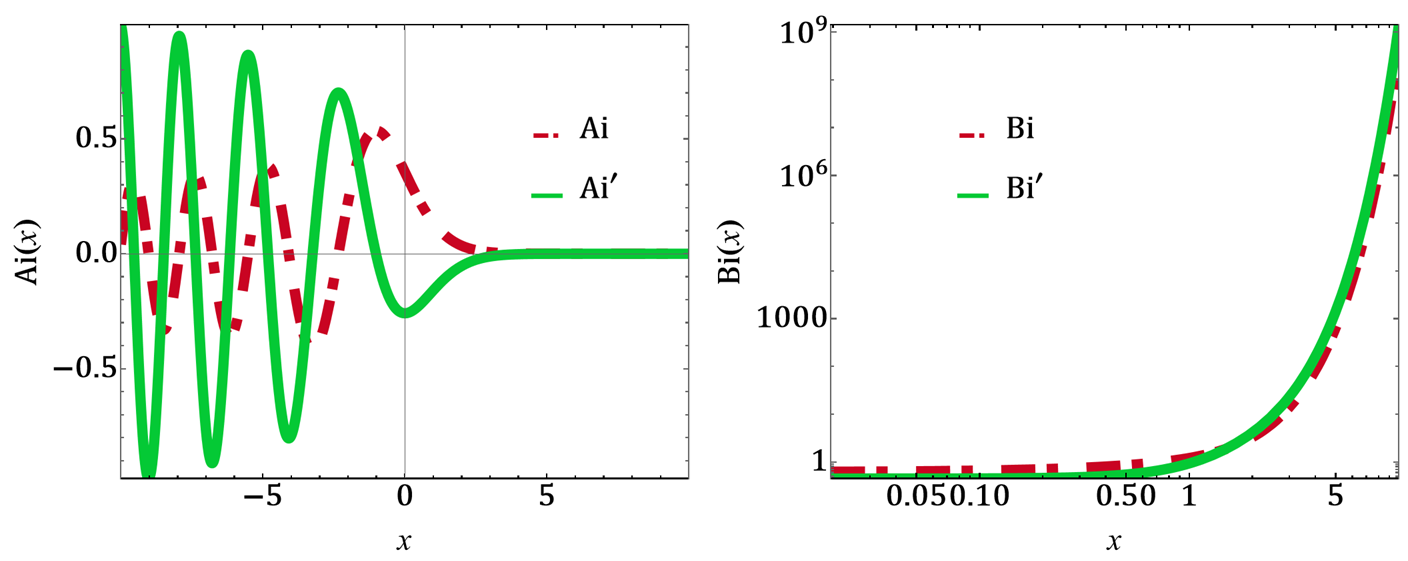}
    \caption{General behaviour of the Airy functions and its derivatives. Note that the panel for the Bi function is in log-log scale.}
    \label{App1}
\end{figure}

Expansions and power laws for the Airy functions can be found for example in \cite{abramowitz}. The general behaviour of these functions is illustrated in Fig. (\ref{App1})

\subsection{D-Parabolic Cylinder functions}

The parabolic cylinder functions originate from the differential equation
\begin{equation}
    \dfrac{d^2f}{dz^2}(z)+(az^2+bz+c)f(z)=0\,,
\end{equation}

\noindent which surges naturally from Laplace's equation when we separate its variables in parabolic cylindrical coordinates.

A particular case of this equation is the so-called Weber equation,
\begin{equation}
    \dfrac{d^2f}{dz^2}(z)+\left(a-\dfrac{1}{4}z^2\right)f(z)=0\,.
\end{equation}

The independent solutions of this equation are commonly expressed as \cite{abramowitz}
\begin{align}
y_1(a; z) &= \exp(-z^2 / 4) \, {}_1F_1\left( \tfrac{1}{2}a + \tfrac{1}{4}; \tfrac{1}{2}; \tfrac{z^2}{2} \right)\,,\\
y_2(a; z) &= z \exp(-z^2 / 4) \, {}_1F_1\left( \tfrac{1}{2}a + \tfrac{3}{4}; \tfrac{3}{2}; \tfrac{z^2}{2} \right)\,,
\end{align}

\noindent where ${}_1F_1(a; b; z)$ is the confluent hypergeometric function. 

The parabolic cylinder functions are particular combinations of the previous equations, defined as
\begin{align}
U(a, z) &= \frac{1}{2^x \sqrt{\pi}} \left[ \cos(x\pi)\Gamma\left(\tfrac{1}{2} - x\right) y_1(a, z) - \sqrt{2} \sin(x\pi)\Gamma(1 - x)\, y_2(a, z) \right]\,,\\
V(a, z) &= \frac{1}{2^x \sqrt{\pi} \Gamma(1/2 - a)} \left[ \sin(x\pi)\Gamma\left(\tfrac{1}{2} - x\right) y_1(a, z) + \sqrt{2} \cos(x\pi)\Gamma(1 - x)\, y_2(a, z) \right]
\end{align}

\noindent where $x = \tfrac{1}{2}a + \tfrac{1}{4}$. 

In this paper, we are concerned about the particular case of the Weber equation
\begin{equation}
        \dfrac{d^2\psi}{dz^2}(z)+\left(\sigma+\dfrac{1}{2}-\dfrac{1}{4}z^2\right)\psi(z)=0\,,
\end{equation}

\noindent which is Schr\"odinger's equation for the harmonic oscillator when we identify $x=z\sqrt{\hbar/2m\omega}$ and $\sigma=\dfrac{\mathcal{E}}{\hbar\omega}-\dfrac{1}{2}$. In this case, following Whittaker \cite{whittaker1990}, the solutions are given in terms of the $D-$parabolic cylinder functions as given in Eq. \ref{solOH}, such that
\begin{align}
U(a, z) &= D_{-a - \frac{1}{2}}(z)\,,\\
V(a, z) &= \frac{\Gamma\left( \tfrac{1}{2} + a \right)}{\pi} \left[ \sin(\pi a) D_{-a - \frac{1}{2}}(z) + D_{-a - \frac{1}{2}}(-z) \right]\,.
\end{align}

The advantage to write the $D-$parabolic cylinder function in terms of $U(a,z)$ is that it is relatively easy to expand $U(a,z)$ in a power series of $z$,
\begin{equation}
U(a, z) = \frac{\sqrt{\pi} \, 2^{- \frac{a}{2} - \frac{1}{4}}}{\Gamma\left( \frac{a}{2} + \frac{3}{4} \right)}
- \frac{\sqrt{\pi} \, 2^{- \frac{a}{2} + \frac{1}{4}}}{\Gamma\left( \frac{a}{2} + \frac{1}{4} \right)} z
+ \frac{\sqrt{\pi} \, 2^{- \frac{a}{2} - \frac{5}{4}}}{\Gamma\left( \frac{a}{2} + \frac{3}{4} \right)} z^2
- \cdots\,,
\end{equation}

\noindent such that 
\begin{equation}
    U(a,0)=\dfrac{\sqrt{\pi}2^{-a/2-1/4}}{\Gamma\left(\dfrac{a}{2}+\dfrac{3}{4}\right)}\quad\text{and}\quad U'(a,0)=-\dfrac{\sqrt{\pi}2^{-a/2+1/4}}{\Gamma\left(\dfrac{a}{2}+\dfrac{1}{4}\right)}\,.
\end{equation}

In this way,
\begin{equation}
      D_{\frac{E}{\nu}-\frac{1}{2}}(0)=\dfrac{\sqrt{\pi}2^{E/2\nu-1/4}}{\Gamma\left(-\dfrac{E}{2\nu}+\dfrac{3}{4}\right)}\quad\text{and}\quad D'_{\frac{E}{\nu}-\frac{1}{2}}(0)=-\dfrac{\sqrt{\pi}2^{E/2\nu+1/4}}{\Gamma\left(-\dfrac{E}{2\nu}+\dfrac{1}{4}\right)}\,.
\end{equation}

Taking $\nu=1$ we get the expressions used in Sec. \ref{sec3}.

Finally, a useful relation is that, when $\sigma\in\mathds{N}$, $D_\sigma(z)$ becomes proportional to a Hermite polynomial \cite{bateman2006higher}:
\begin{equation}
    D_\sigma(z)=\mathrm{e}^{-z^2/4}2^{-\sigma/2}H_\sigma(z/\sqrt{2})\,.
\end{equation}

As the not sheared harmonic oscillator is such a case with $\sigma \in \mathds{N}$, the route of solving the problem with $D-$parabolic cylinder functions is almost never taken. In fact, almost all textbooks do not even mention this function (Merzbacher \cite{merzbacher1998quantum} uses it but just in the context of a double harmonic oscillator).

\begin{figure}[t]
    \centering
    \includegraphics[width=0.75\linewidth]{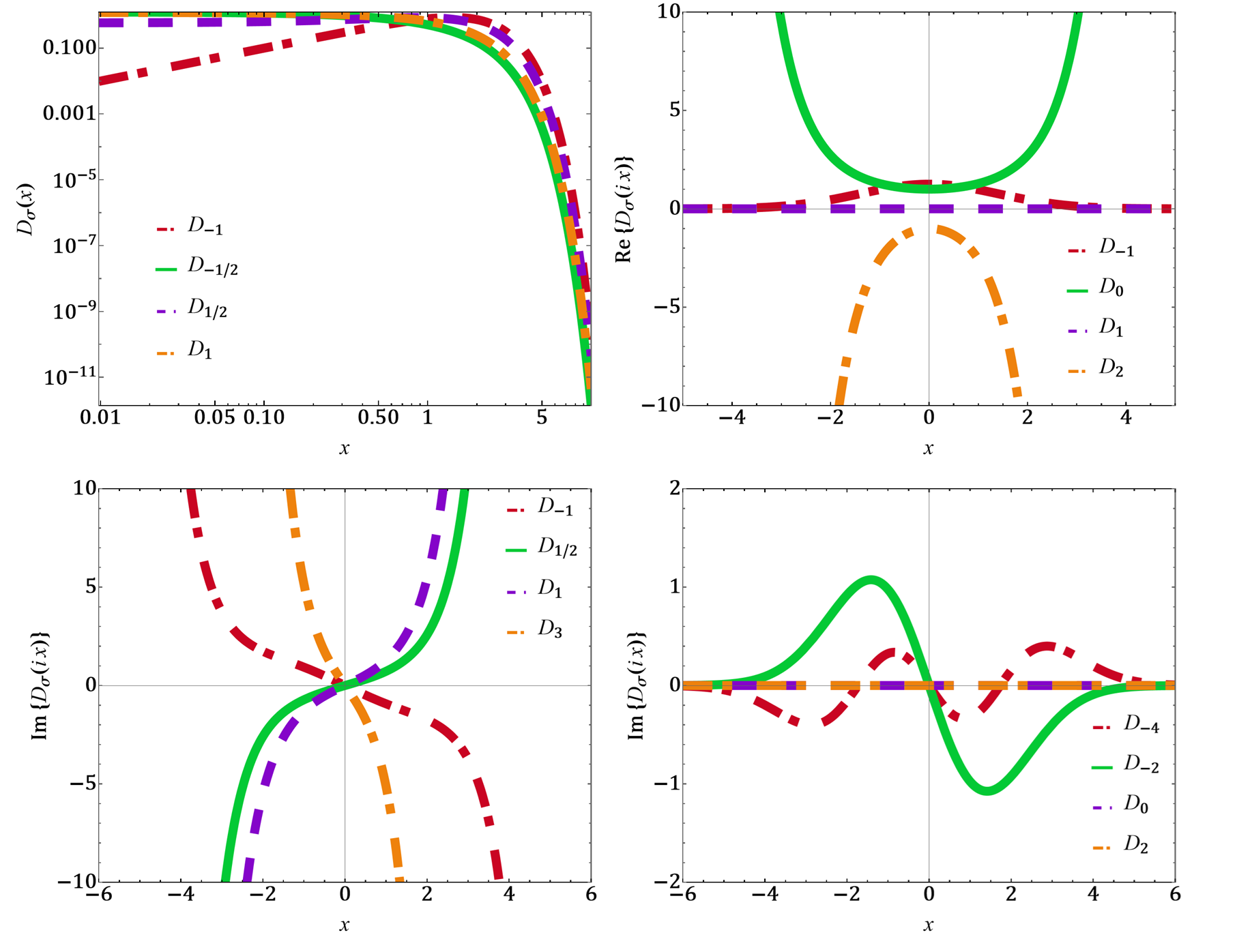}
    \caption{General behaviour of the $D-$parabolic function for real and imaginary arguments. For real arguments, the function is purely real  and falls out rapidly (left-superior panel). For imaginary arguments, the imaginary part  converges only if  $\sigma$ is even (bottom panels), and the real part  converges only if $\sigma$ is odd (right-superior panel).}
    \label{App2}
\end{figure}

Expansions and power laws for the $D-$parabolic cylinder function can be found for example in \cite{KMAbadir1993}. The general behaviour of $D_\sigma(x)$ is illustrated in Fig. \ref{App2}.

%\printbibliography

\end{document}